\documentclass[journal,draftcls,onecolumn,12pt,twoside]{IEEEtranTCOM}

\usepackage[]{fontenc}
\usepackage{color}
\usepackage{cite}
\usepackage[pdftex]{graphicx}
\usepackage{amsmath}

\normalsize

\usepackage[cmintegrals]{newtxmath}
\usepackage{algorithm}
\usepackage{algorithmic}
\usepackage{array}
\usepackage{multirow}
\ifCLASSOPTIONcompsoc
\usepackage[caption=false,font=normalsize,labelfont=sf,textfont=sf]{subfig}
\else
\usepackage[caption=false,font=footnotesize]{subfig}
\fi

\usepackage{stfloats}
\usepackage{tikz}

\begin{document}

\title{Optimal Policies for Status Update Generation in a Wireless System with Heterogeneous Traffic}

\author{George~Stamatakis, Nikolaos Pappas~\IEEEmembership{Member,~IEEE},~and~Apostolos~Traganitis
\thanks{G.~Stamatakis and A.~Traganitis are with the
Institute of Computer Science, Foundation for Research and Technology - Hellas (FORTH) email:\{gstam,tragani\}@ics.forth.gr.}
\thanks{N.~Pappas is with the Department of Science and Technology, Link\"{o}ping University, Norrk\"{o}ping SE-60174, Sweden email:nikolaos.pappas@liu.se.}
\thanks{This work was supported in part by the Center for Industrial Information Technology (CENIIT), ELLIIT, and the EU project DECADE under Grant H2020-MSCA-2014-RISE: 645705, the European Union’s Horizon 2020 research and innovation programme.}}

\maketitle
\vspace{-0.5in}
\begin{abstract}
A large body of applications that involve monitoring, decision making, and forecasting require timely status updates for their efficient operation. 
Age of Information (AoI) is a newly proposed metric that effectively captures this requirement. Recent research on the subject has derived AoI optimal policies for the generation of status updates and AoI optimal packet queueing disciplines.
Unlike previous research we focus on low-end devices that typically support monitoring applications in the context of the Internet of Things.
We acknowledge that these devices host a diverse set of applications some of which are AoI sensitive while others are not. 
Furthermore, due to their limited computational resources they typically utilize a simple First-In First-Out (FIFO) queueing discipline.
We consider the problem of optimally controlling the status update generation process for a system with a source-destination pair that communicates via a wireless link, whereby the source node is comprised of a FIFO queue and two applications, one that is AoI sensitive and one that is not.
We formulate this problem as a dynamic programming problem and utilize the framework of Markov Decision Processes to derive optimal policies for the generation of status update packets.
Due to the lack of comparable methods in the literature, we compare the derived optimal policies against baseline policies, such as the zero-wait policy, and investigate the performance of all policies for a variety of network configurations.
Results indicate that existing status update policies fail to capture the trade-off between frequent generation of status updates and queueing delay and thus perform poorly.
\end{abstract}

\IEEEpeerreviewmaketitle

\section{Introduction}
\label{sec:Introduction}

Applications that offer monitoring, informed decision making, and forecasting services in cyber-physical systems, often rely on timely status updates~\cite{stankovic2014research}.
A large number of such applications has been developed in the context of Internet of Things with examples that include, but are not limited to, smart cities, smart factories and grids, smart agriculture, parking and traffic management, water management, e-Health, environment monitoring and education~\cite{stankovic2014research, tragos2014enabling}.
The proliferation of these applications is expected to have a profound impact on key sectors of economy, and this has spurred research on their particular operational requirements~\cite{LMT2014}. 
A key result in the field was the realization, by the authors in~\cite{kaul2012real}, that the objective of timely status updating is not captured by metrics such as utilization and delay, which are typically used in network design and management.
To alleviate this problem, a new metric called Age of Information (AoI) was introduced in~\cite{kaul2012real} to effectively capture the requirement for timely updating.

Recent works~\cite{kaul2011minimizing, bedewy2017age} have shown that minimization of the AoI, for a status update system, can be achieved by optimally choosing the generation rate of packets that carry status update information as well as by changing the queuing discipline from First In First Out (FIFO) to Last In First Out (LIFO).
However, considering the complexity of monitoring, decision making and forecasting we expect that the deployed network equipment will support a large number of applications with diverse network requirements, e.g., some of them will be AoI-sensitive while others will not. 
Furthermore, we expect that a LIFO service policy by the queue might not be compatible with the network requirements of other applications.
For example, management and software update as well as video transmission applications typically require a FIFO queue.
Finally, in the case of status update applications it is customarily assumed that the wireless device has some control over the packet generation rate, however, this may not be true for all the applications it supports. 

In this work, we consider a discrete time wireless system with a single source-destination pair where 
the source node is comprised of a wireless transmitter and a single queue which admits both status and non-status update packets. 
Status update packets are generated by a process that is fully controlled by the source node while non-status update packets are generated by an application which is beyond the control of the source node.
All transmissions are subject to failure and upon a failed transmission attempt, the head-of-line packet will be retransmitted up to a maximum number of times after which it will be dropped.
\textit{Furthermore, we assume that the AoI of the system is constrained to always be less than a predefined threshold value. 
In case this constraint is not satisfied the source node will change temporarily the queue's default FIFO service policy, and its transmission scheme so that the delivery of a fresh status update to the destination is guaranteed and all outdated status update packets in the queue are dropped}.

To the best of our knowledge, this is the first work to consider the design of an optimal controller for the generation of status updates for the wireless system under consideration.
Optimality here is taken with respect to a cost function that is additive over time and depends on both the AoI of the system and the cost related to the use of the mechanism that guarantees a successful packet transmission. 
We formulate the problem at hand as a dynamic programming problem and utilize the framework of Markov Decision Processes (MDP) to derive optimal policies.
Finally, we show by comparison that for a wide range of scenarios well known policies from the literature, such as the zero-wait policy, perform poorly for the system under consideration.

The remainder of the paper is organized as follows. 
In Section~\ref{sec:RelatedWork} we present recent work related to the problem described above. 
In Section~\ref{sec:SystemModel} we present the system model considered in this work.
In Section~\ref{sec:ProblemFormulation} we formulate this problem as a dynamic programming problem.
In Section~\ref{sec:AgeOptimalPolicy} we show that the dynamic program constitutes a MDP and present the algorithms we use to derive the AoI optimal policies.
Finally, in Section~\ref{sec:Results} we present numerical results for the evaluation of the derived policies.
Our conclusions are in Section~\ref{sec:conclusion}.

\section{Related Work}
\label{sec:RelatedWork}

In this section we present related work divided in two categories. 
The first category includes works that follow a queueing theoretic approach to the performance analysis and optimization of communication systems with respect to AoI and AoI related metrics while the second category includes works that focus on scheduling with respect to AoI.

In \cite{HuangISIT2015} the AoI in a general multi-class $M/G/1$ queueing system is studied. 
In addition, the exact peak-age-of-information (PAoI) expressions for both $M/G/1$ and $M/G/1/1$ systems are obtained. 
The work in \cite{KamTIT2016} studied the status age of update packets transmitted through a network. 
The authors modeled a network as an $M/M/ \infty$ model, and they derived the expression for the average AoI. 
The PAoI in an $M/M/1$ queueing system with packet delivery errors is considered in \cite{ChenISIT2016}.

The work in \cite{Yates2016arxivSHS} considers multiple independent sources that transmit status updates to a monitor through simple queues. A new simplified technique for evaluating the AoI in finite-state continuous-time queueing systems is derived. The technique is based on stochastic hybrid systems and makes AoI evaluation to be comparable in complexity to finding the stationary distribution of a finite-state Markov chain. In \cite{InoueISIT2017} the stationary distributions of AoI and the PAoI are considered.
The authors derived  explicit formulas for the Laplace-Stieltjes transforms of the stationary distributions of the AoI and PAoI in FCFS $M/GI/1$ and $GI/M/1$ queues. Yates in \cite{Yatesarxiv201806} employed stochastic hybrid systems to enable evaluation of all moments of the age as well as the moment generating function of the age in any network that can be described by a finite-state continuous-time Markov chain.

In \cite{kosta2017age}, the authors introduce the metrics of Cost of Update Delay (CoUD) and Value of Information of Update (VoIU) in order to characterize the cost of having stale information at a remote destination and to capture the reduction of CoUD upon reception of an update respectively.
The work in \cite{SunTIT2017} studied the optimal control of status updates from a source to a remote monitor. 
The authors showed that in some cases, the optimal policy is to wait for a certain time before submitting a new update. 
In \cite{KostaGlobecom2018} the authors study the average AoI of a primary node and the throughput of multiple secondary nodes in a shared access network with priorities.
In \cite{kam2018packet}, the average AoI for an $M/M/1/2$ queueing system with packet deadlines is studied.

Next we present works that focus on scheduling.
The work in \cite{KadotaAllerton2016} considers a wireless broadcast network with a base station sending time-sensitive information to a number of nodes. 
A discrete-time decision problem is formulated to find a scheduling policy that minimizes the expected weighted sum of AoI for all nodes in the network.
The authors in \cite{NajmISIT2018} consider a stream of status updates where each update is either of high priority or an ordinary one. 
Then, a transmission policy that treats updates depending on their priority is considered. 
The arrival processes of the two kinds of updates are modeled as independent Poisson processes while the service times are modeled as two exponentials.
The work in \cite{SunISIT2017} considers a problem of sampling a Wiener process, the samples are forwarded to a remote estimator via a channel that consists of a queue with random delay. The estimator reconstructs a real-time estimate of the signal. The optimal sampling strategy that minimizes the mean square estimation error subject to a sampling frequency constraint is studied.

In \cite{Talakarxiv201803} the problem of AoI minimization for single-hop flows in a wireless network, under interference constraints and a time varying channel is considered. 
A class of distributed scheduling policies, where a transmission is attempted over each link with a certain attempt probability is studied.
AoI minimization for a network under general interference constraints and a time varying channel is studied in \cite{TalakWiOpt2018} and~\cite{TalakISIT2018} with known and unknown channel statistics respectively.
The work in \cite{LuMobihoc2018} proposed a real-time algorithm for scheduling traffic with hard deadlines that provides guarantees on both throughput and AoI. 
The work in \cite{QingTIT2018} considered a set of transmitters, where each transmitter contains a given number of status packets and all share a common channel. 
The problem of scheduling transmissions in order to minimize the overall AoI is considered.
The authors in \cite{SunElifInfWork2018} study an AoI minimization problem, where multiple flows of update packets are sent over multiple servers to their destinations. 
The authors in \cite{KamInfwork2018}, considered an alternative metric, the effective age, in order to achieve lower estimation error in a remote estimation problem. 
The problem they considered for developing an effective age is the remote estimation of a Markov source.

The work in~\cite{ACC2016} considers a sequential estimation and sensor scheduling problem in the presence of multiple communication channels. 
In \cite{WCNC2018}, scheduling the transmission of status updates over an error-prone communication channel is studied in order to minimize the average AoI at the destination under a constraint on the average number of transmissions at the source node. 
The work in \cite{SIU2018Elif}, introduced a deep reinforcement learning-based approach that can learn to minimize the AoI with no prior assumptions about network topology. 

Additional references can be found in the survey \cite{NOW-AoI}.

\section{System Model}
\label{sec:SystemModel}

\begin{figure}[!htb]
	\centering
	\includegraphics[scale=0.7]{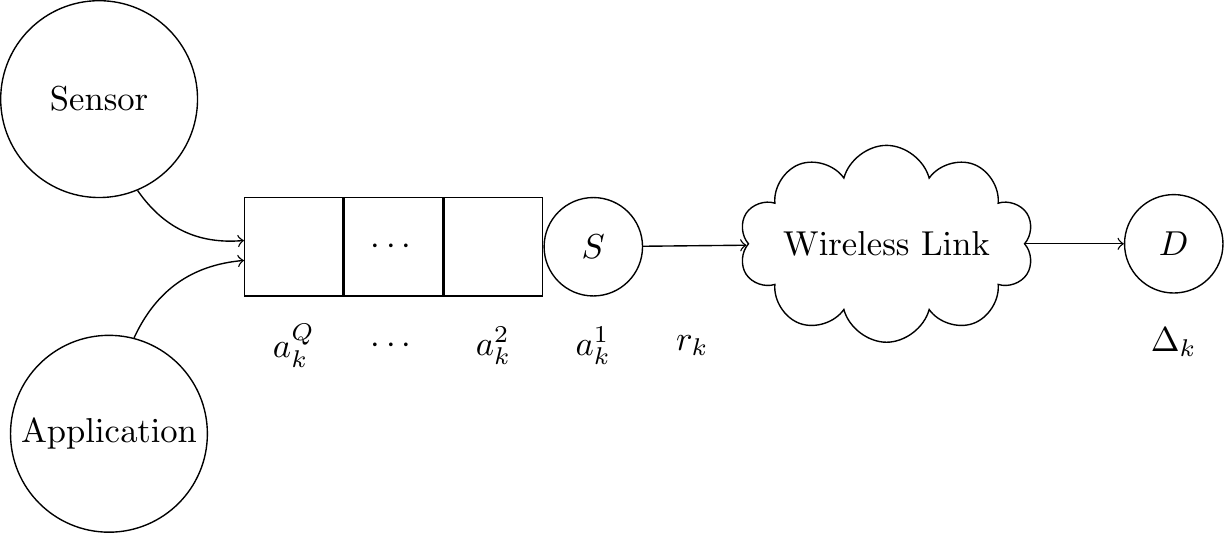}
	\vspace{-0.2in}
	\caption{System model.}
	\label{fig:systemModel}
\end{figure}

We consider the system depicted in Fig.~\ref{fig:systemModel}, which is comprised of a source node that transmits data to a destination node $D$ through a wireless link. 
The source node consists of a sensor that generates data packets with status update information, an application that generates data packets with \emph{non-}status update information, a finite queue, and a transmitter $S$.
Subsequently, we will use the term \emph{status updates} to refer to packets conveying status update information and the term \emph{application packets} to refer to packets with non-status update information. 

We assume that time is slotted and a single packet is transmitted in each time-slot.
At the beginning of the $k$-th time-slot, $S$ will commence the transmission of the head of line packet. 
The transmission may succeed with probability $P_s$ or fail with probability $1-P_s$, independently of the transmission outcomes in previous time-slots.
We assume that all packet transmissions are acknowledged so that the success or failure of the transmission will be known to the source node by the end of the $k$-th time-slot.
In the case of a failed transmission, a retransmission counter $r_k$ will be incremented and the packet will be retransmitted during the next time-slot.
The server will make up to $r_{max}$ transmission attempts before dropping the packet and proceeding with the next one in queue. 

Within the duration of a time-slot, the application in Fig.~\ref{fig:systemModel} will generate a single application packet per time-slot with probability $P_a$, while the source node, which is in full control of the sensor, has to \emph{decide} whether to generate a fresh status update or not. 
All packets generated within the duration of a time-slot will be enqueued unless the queue is full, in which case they will be dropped. 

Finally, we assume that the source node must satisfy a hard constraint on AoI, i.e., $\Delta_k$ should always be less than a threshold value $\Delta_{\max}$. 
In case $\Delta_k$ becomes equal to $\Delta_{\max}$, the queue's service policy will change temporarily from its default FIFO operation so that the source node may be able to apply the following three actions:
\begin{enumerate}
\item The head of line packet is dropped.
\item All status update packets currently in the queue are dropped.
\item A fresh status update packet is sampled and transmitted with success probability 1 by the transmitter.
\end{enumerate}
\textit{We emphasize that this type of transmission induces a high penalty, due to the cost of the mechanism that guarantees a successful transmission, and that it is available to the source node only when AoI reaches the threshold value}.
Further details regarding the conditions that determine the set of available decisions to the source node will be presented in Section~\ref{sec:ProblemFormulation}.

Our objective is to devise a controller that generates status updates so as to minimize the expected value of a cost function which is additive over time, i.e., at the end of each time-slot a new cost value is added to the aggregate cost of all previous time-slots. 
This new cost value will be either equal to the AoI or equal to a fixed value which is much larger than $\Delta_{\max}$, in case the mechanism that guarantees a successful transmission was used.
In this work we consider the problem of minimizing the expected value of the additive cost function over an infinite time horizon. 
To guarantee that the produced infinite sum converges, we utilize discounting, i.e., the importance of future costs reduces with time.

\section{Problem Formulation}
\label{sec:ProblemFormulation}
In this section we formulate a dynamic programming problem for the system considered. We begin with the description of the state, control and random variable spaces and proceed with the system transition function, the state transition costs and the system cost function, which is additive over time. Finally, we give a short description about the optimal policy and its characteristics.
\subsubsection*{State Space Description} We utilize the AoI metric, denoted with $\Delta_k,\ k = 0, 1, \dots$ in Fig.~\ref{fig:systemModel}, to characterize the freshness of status updates at destination $D$. 
AoI was defined in \cite{kaul2012real} as the time that has elapsed since the generation of the last status update that has been received by $D$.
More specifically, let $\tau_m$ denote the generation time-stamp of the $m$-th status update, $\tau_m'$ be the time-slot that the $m$-th status update arrived at destination $D$ and $M_k$ be the index of the last status update that $D$ has received by the $k$-th time-slot, i.e., $M_k = \max \{m | \tau_m' \leq k \}$, then $\Delta_k = k - \tau_{M_k}$.
This representation of $\Delta_k$, for a centralized control model, has the drawback that one must keep time-stamps as part of the description of the system's state which can be computationally inefficient.

\begin{figure}[!h]
	\centering
	\includegraphics[scale=0.65]{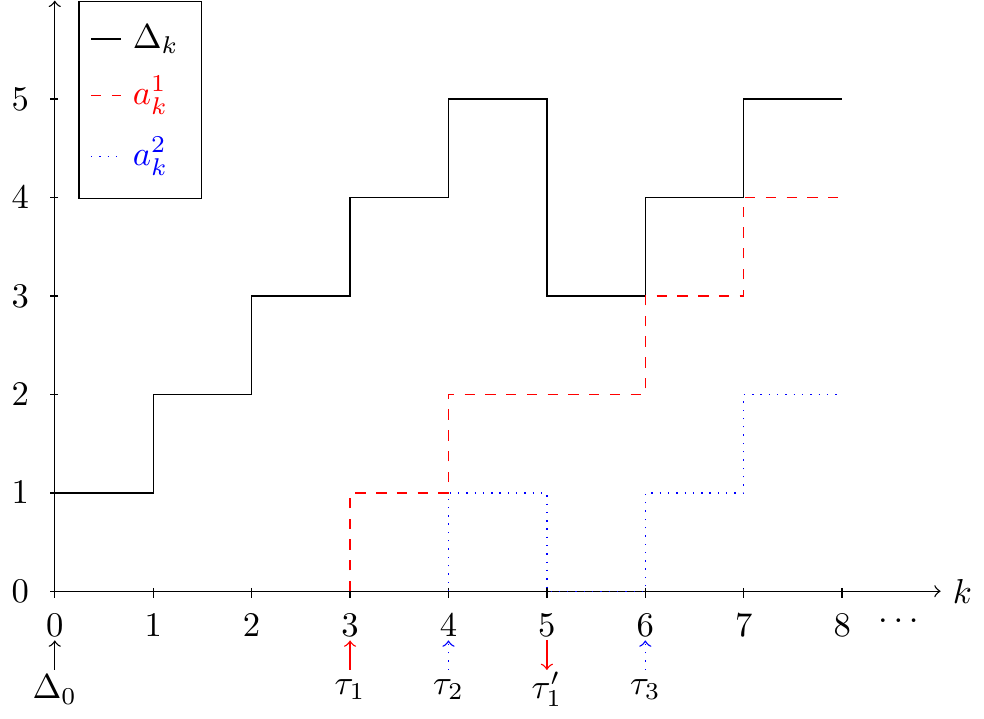}
	\vspace{-0.2in}
	\caption{Evolution of $\Delta_k$, $a_k^1$, and $a_k^2$ over time for an example scenario with three status update arrivals at $\tau_1$, $\tau_2$, and $\tau_3$ and one departure at $\tau_1'$ when $\Delta_0$ is set to one.}
	\label{fig:agesOverTime}
\end{figure}

We can derive a different expression for $\Delta_k$ by noting that when $k$ equals the time-slot of the last status update arrival, i.e., $k = \tau_{M_k}'$, we have $\Delta_{\tau_{M_k}'} = \tau_{M_k}' - \tau_{M_k} + 1$, where the increment by one is due to the slotted time assumption whereby we account for the next time-slot in advance.  
For the system of Fig.~\ref{fig:systemModel} the time interval, $\tau_{M_k}' - \tau_{M_k}$, is equal to the total time that the $M_k$-th status update spent waiting in queue and under service by the transmitter. 
We define $a_{M_k} = \tau_{M_k}' - \tau_{M_k}$ and thus $\Delta_{\tau_{M_k}'} = a_{M_k} + 1$.
The AoI will increase by one unit for each time-slot that passes by beyond $\tau_{M_k}'$, 
i.e., $\Delta_k = a_{M_k} + 1 + \delta_k,\ k \geq \tau_{M_k}'$, where $\delta_k = k - \tau_{M_k}'$.
For the source node to have knowledge of $a_{M_k}$ for each status update delivered to $D$, we associate with each queue position a counter $a_k^{q},\ q=2, \dots, Q$ (see Fig.~\ref{fig:systemModel}) that holds the total time spent waiting in queue for the status update currently occupying the queue position. 
For the status update currently under service, counter $a_k^1$ holds the aggregate time it has spent waiting in the queue and under service. 
The process of updating the values of $a_k^q,\ q=1, 2,\dots, Q$ as packets move from one queue position to the next will be presented later in this section.
Fig.~\ref{fig:agesOverTime} depicts the evolution of $\Delta_k$, $a_k^1$ and $a_k^2$ over time for an example scenario.
Furthermore, delay information for the application packets is irrelevant for the AoI of the system, thus whenever an application packet occupies the $q$-th position of the queue we assign counter $a_k^q$ the special value of $-1$.
An additional advantage of this assignment is a significant reduction in the size of the state space. 
Finally, we assume that whenever $\Delta_k$ becomes equal to $\Delta_{\max} \in \mathbb{Z}^+$ the source node will preemptively drop the packet currently under service along with all queued status updates and it will transmit a fresh status update through an error free but expensive channel.
As a consequence of this assumption, $\Delta_k$ will be bounded above by $\Delta_{\max}$, and the state space will be finite, as will become apparent subsequently. 

The state of the system of Fig.~\ref{fig:systemModel}, at the beginning of the $k$-th time slot, is determined by the values of $\Delta_k$, $r_k$ and $a_k^q, q=1, 2,\dots, Q$.
We use column vector $x_k = [\Delta_k, r_k, a_k^1, \dots, a_k^Q]^T$ to refer collectively to the state of the system and $X$ to denote the set of all system states. 

\subsubsection*{Control/Decision Space Description} 
Given its current state $x_k$ the source node has to make a decision regarding the generation of a fresh status update, the dropping of the head of line packet and the preemptive transmission of a fresh status update.
We denote the set of all possible decisions, termed controls, with, 
\begin{equation}
U = \{(u^s, u^d, u^{p}): u^s, u^d, u^{p} \in \{0,1\},\, \neg u^{p} \lor (u^{p} \land u^s \land u^d)\}
\end{equation}
where $u^s$ is a binary variable indicating whether the sensor should generate a status update, $u^d$ is a binary variable indicating whether the head of line packet should be dropped at the end of the current time-slot in case of another failed transmission. $u^{p}$ is a binary variable indicating whether all status updates within the source node should be dropped and a fresh status update should be preemptively transmitted to $D$. 
Predicate $\neg u^{p} \lor (u^{p} \land u^s \land u^d)$ will evaluate to true either for $u^{p} = 0$ ($\neg u^p = 1$) along with all combinations $(u^s, u^d)\in\{0,1\}^2$, or for $(u^s, u^d, u^{p}) = (1,1,1)$.
The latter control involves generating a fresh status update ($u^s=1$), dropping the head of line packet ($u^d=1$) as well as dropping all queued status updates and preemptively transmitting the fresh status update by using the costly, yet error free channel ($u^p = 1$).

At each system state $x$ only a subset of the controls in $U$ will be available to the source node.
This subset is typically called the constraint control set and is denoted with $U(x) \subseteq U$.
Table~\ref{tbl:constraintControlSet} categorizes the states based on their attributes and presents the corresponding constraint control sets. 
For notational convenience we drop the time index $k$ since constraint control sets do not change over time.

\begin{table*}[!t]
	\centering
	\caption{Set of available controls at different state subsets.}
	\vspace{-0.2in}
	\label{tbl:constraintControlSet}
	\begin{tabular}{|p{0.3\textwidth} | m{0.2\textwidth}| m{0.45\textwidth}|}
		\hline
		Subset of States   & Constraint Control Set $U(x)$   & Description \\
		\hline 
		$\{x \in X:\Delta = \Delta_{\max}\}$ & $\{(1,1,1)\}$ & Generate a fresh status update, drop the head of line packet and all queued status updates, preemptively transmit a fresh status update through the expensive channel.\\
		\hline  
		$\{x \in X: \Delta \neq \Delta_{\max}, r = r_{\max}, a^Q \neq 0\}$  & $\{ (0,1,0) \}$  & The source node will not generate a fresh status update due to the full queue ($a^Q \neq 0$), the head of line packet will be dropped at the end of the current time-slot in case the $r_{\max}$-th retransmission fails. \\
		\hline 
		$\{x\in X: \Delta \neq \Delta_{\max}, r = r_{\max}, a^Q = 0\}$  & $\{(0, 1, 0), (1, 1, 0)\}$ & The source node may or may not generate a fresh status update while the head of line packet will be dropped in case the $r_{\max}$-th retransmission fails. \\
		\hline 
		$\{x\in X: \Delta \neq \Delta_{\max}, r \neq r_{\max}, a^Q \neq 0\}$  & $\{(0, 0, 0)\}$ & The sensor cannot generate a fresh status update due to the queue being full.\\
		\hline
		Otherwise &  $\{(0, 0, 0), (1, 0, 0)\}$ &  The source may or may not generate a fresh status update.\\
		\hline
	\end{tabular}
\end{table*}

\subsubsection*{System Random Variables} At the beginning of the $(k+1)$-th time-slot the system will make a transition to a new state $x_{k+1}$ as a result of the selected control $u_k$ and two random events.
The first one is the arrival of an application packet which is represented by the binary random variable $W_k^a$ and the second one is the successful transmission of the head-of-line packet  which is represented by the binary random variable $W_k^s$. 
As mentioned in section~\ref{sec:SystemModel} we assume that the application in Fig.~\ref{fig:systemModel} will generate a single packet per time-slot with probability $P_a$.
Furthermore, the transmitter will deliver a packet successfully with probability $P_s$ independently of the transmission outcome in any previous time-slot.
The probability distributions of $W_k^a$ and $W_k^s$ are assumed to be independent of previous time-slots and identically distributed for all time-slots.
We use the random vector $W_k = [W_k^a, W_k^s]^T$ to collectively refer to the random variables of the system.

\subsubsection*{State Transition Function} Given $x_k$, $u_k$ and the values for $W_k^s$ and $W_k^a$, which will be known to the source node by the end of the $k$-th time-slot, the system will make a transition to a new state $x_{k+1} = [\Delta_{k+1}, r_{k+1}, a_{k+1}^1, \dots, a_{k+1}^Q]^T$.
This transition is determined by the discrete-time system $x_{k+1} = f(x_k, u_k, W_k)$.
Next we present the elements that comprise $f(\cdot)$.	
We begin with $\Delta_{k+1}$ which is given by the following expression,
\begin{equation}
\label{eq:updateAoI}
	\Delta_{k+1} = \begin{cases}
		1, & \mbox{if } x_k \in X_{\Delta_{\max}}\\
		\Delta_k +1, & \mbox{if } x_k \not\in X_{\Delta_{\max}} \text{ and } (W_k^s = 0 \text{ or } a_k^1 = -1) \\
		a_k^1 + 1, & \mbox{if } x_k \not\in X_{\Delta_{\max}}\text{ and } W_k^s =1\text{ and } a_k^1 \neq -1,			
	\end{cases}
\end{equation}
where $X_{\Delta_{\max}} = \{x \in X:  \Delta = \Delta_{\max} \}$. 
Expression~(\ref{eq:updateAoI}) shows that $\Delta_{k+1}$ will be set to one whenever AoI becomes equal to the maximum acceptable value of $\Delta_{\max}$. This is due to the transmission of a fresh status update through an error free channel.
Furthermore, from (\ref{eq:updateAoI}) we see that $\Delta_k$ will be incremented by one in the cases of an unsuccessful packet transmission and that of a successful transmission of an application packet. 
Finally, in the case of a successful transmission of a status update, $\Delta_{k+1}$ will be set to $a_k^1+1$ which is equal to $(\tau_{M_k}' - \tau_{M_k}) + 1$.

Assuming that the queue in Fig.~\ref{fig:systemModel} can store at least one more packet besides the one currently under service, i.e., $Q~>~1$, the value of the retransmission counter $r_k$ is updated as follows,
\begin{equation}
\label{eq:retransmissionCounter}
r_{k+1} = \begin{cases}
	0, & \text{if } a_k^1 = 0 \text{ and } u_k^s = w_k^a = 0\\
	0, &  \text{if } (w_k^s = 1 \text{ or } u_k^d = 1) \text{ and } a_k^2 = u_k^s = w_k^a = 0\\ 
	1, & \text{if } (u_k^d = 1 \text{ or } w_k^s = 1 ) \text{ and } (a_k^2 \neq 0 \text{ or } u_k^s = 1 \text{ or } w_k^a = 1) \\
	r_k + 1, &  \text{if } a_k^1 \neq 0 \text{ and } w_k^s = 0 \text{ and } u_k^d = 0. 
\end{cases}
\end{equation}
From (\ref{eq:retransmissionCounter}) we see that $r_{k+1}$ will be set to zero when there is no packet for the transmitter to transmit at the beginning of the $(k+1)$-th time-slot. This may occur in two cases.
Firstly, in case there wasn't a packet under service ($a_k^1 = 0$) and, additionally, there were no packet arrivals $(u_k^s = w_k^a = 0)$ during the $k$-th time-slot.
Secondly, in case the packet under service was either successfully transmitted or dropped $(w_k^s = 1 \text{ or } u_k^d = 1)$, the queue was empty ($a_k^2=0$ implies that all queue positions with $q \geq 2$ were also empty) and there were no packet arrivals $(a_k^2 = u_k^s = w_k^a = 0)$ during the $k$-th time-slot.

On the other hand, $r_{k+1}$ will be set to one if the packet being transmitted at the $k$-th time-slot departed from the source node either by being successfully transmitted or by being dropped and there exists another packet for the transmitter to transmit at the beginning of the $(k+1)$-th time-slot.
This scenario will occur either if the queue position with $q=2$ was occupied by a packet during the $k$-th time-slot, i.e., $a_k^2 \neq 0$, or in case it was empty and a new packet arrived at the source node during the $k$-th time-slot.
Finally, the value of $r_{k+1}$ will be incremented by one if there exists a packet under service $a_k^1 \neq 0$ which is neither transmitted successfully, nor is it dropped by the source node.

Now, let $N_k^m \in \{0,\dots,Q\}$ be zero, in case the queue is empty, and equal to the index value $q$, of the last queue position which is occupied by a packet,
\begin{equation}
N_k^m = \begin{cases}
	0, \qquad \text{if } \{q\in 1,\dots, Q\, :\, a_k^q \neq 0\} \text{ is empty}  \\
	\max \{q\in 1,\dots, Q\, :\, a_k^q \neq 0\}, \qquad \text{otherwise}. 
\end{cases}
\end{equation}
Furthermore, let $N_k^p$ denote the number of application packets in queue at the $k$-th time-slot. 

We can distinguish three groups of expressions related to updating the queue delay counter values $a_{k+1}^q,\ q=1,\dots, Q$.
The first group of expressions applies to the case where $x_k \in X_{\Delta_{\max}}$ and is presented in Table~\ref{tbl:delayCountersUpdate1}.
The second group of expressions applies when both $x_k \not\in X_{\Delta_{\max}}$ and the packet that was transmitted at the $k$-th time-slot departed from the system either due to a successful transmission or because it was dropped by the transmitter ($u_k^d = 1 \text{ or } w_k^s = 1$) and is presented in Table~\ref{tbl:delayCountersUpdate2}. 

The third group of equations presented in Table~\ref{tbl:delayCountersUpdate3}, applies in the case where both $x_k \not\in X_{\Delta_{\max}}$ and the packet that was transmitted at the $k$-th time-slot did not depart from the source node which may occur if the packet was neither transmitted successfully nor dropped.

\begin{table*}[!t]
	\centering
	\caption{Update of delay counters when $x_k \in X_{\Delta_{\max}}$.}
	\vspace{-0.2in}
	\label{tbl:delayCountersUpdate1}
	\begin{tabular}{|m{0.1\textwidth} | m{0.42\textwidth}| m{0.4\textwidth}|}
		\hline
		$a_{k+1}^q$  & Conditions for transition  & Description \\
		\hline 
		\hline
		-1 & $q = 1,\dots, N_k^p$ & The first $N_k^p$ queue positions will be occupied exclusively by application packets since all status updates would have been dropped.\\
		\hline
		-1 & $w_k^a = 1\text{ and } q = N_k^p + 1$ & In the case of an application packet arrival, the new packet will be placed in the $(N_k^p+1)$-th queue position, and $a_{k+1}^{N_k^p+1}$ will be set to -1.\\
		\hline
		0 & $w_k^a = 0\text{ and } q = N_k^p + 1$ & In the case of no application packet arrival, $a_{k+1}^{N_k^p+1}$ will be set to zero.\\
		\hline
		0 & $ q = N_k^p + 2, \dots, Q$ & For all remaining queue positions, up to the $Q$-th slot, $a_{k+1}^q$ will be set to zero to indicate that they are empty. \\
		\hline
			\end{tabular}
	\end{table*}

\begin{table*}[]
	\centering
	\caption{Expressions to update delay counters when $x_k \not\in X_{\Delta_{\max}}$ and the head-of-line packet departs.}
	\vspace{-0.2in}
	\label{tbl:delayCountersUpdate2}
	\begin{tabular}{|m{0.1\textwidth} | m{0.42\textwidth}| m{0.4\textwidth}|}
		\hline
		$a_{k+1}^q$  & Conditions for transition  & Description \\
		\hline 
		\hline
		$a_k^{q+1}+1$ &  $a_k^{q+1} > 0 \text{ and } q = 1, \dots, N_k^m-1$ & All packets in the queue will be shifted towards the head-of-line, and, accordingly, the values of $a_k^{q}$ must be shifted to the right, i.e., $a_k^{q+1} \rightarrow a_k^{q}$. Especially for status updates, $a_k^{q+1} > 0$, the corresponding counters $a_k^q$ will be increased by one to indicate that the packets will spend another time-slot in the system.\\
		\hline
		$a_k^{q+1}$   &  $a_k^{q+1} = -1 \text{ and }q = 1, \dots, N_k^m-1$ & Application packets will also be shifted to the right although the values of $a_k^{q}$ will not be incremented by one.\\
		\hline
		$-1$  & $ u_k^s = 0\text{ and } w_k^a = 1 \text{ and } q = N_k^m $ &  Addition of a newly arrived application packet at the first empty queue position.\\
		\hline
		$1$   & $ u_k^s = 1\text{ and } N_k^m < Q \text{ and } q = N_k^m $ & Addition of a new status update at the first empty queue position. Status updates are generated only when the queue is not full ($N_k^m < Q$).\\
		\hline
		$-1$  & $ u_k^s = 1\text{ and } w_k^a = 1\text{ and } N_k^m \leq Q-1, \text{and } q = N_k^m + 1$ & Addition of both a new status update and a new application packet. We assume that status updates enter the queue first. There will always be enough queue slots for both packets given that status updates are generated only if there already exists an empty queue position and, in this case, we also have the departure of the head-of-line packet. \\
		\hline
		$0$   & $ u_k^s = 0\text{ and } w_k^a = 0 \text{ and } q = N_k^m $ & $a_k^q$ counters will be set to $0$ for all empty queue positions.\\
		\hline 
	\end{tabular}
\end{table*}
			
\begin{table*}[]
	\centering
	\caption{Expressions to update delay counters when the head-of-line packet does not depart and $x_k \not\in X_{\Delta_{\max}}$.}
	\label{tbl:delayCountersUpdate3}
	\begin{tabular}{|m{0.1\textwidth} | m{0.42\textwidth}| m{0.4\textwidth}|}
		\hline
		$a_{k+1}^q$  & Conditions for transition  & Description \\
		\hline 
		\hline		
		$a_k^q + 1$ 	& $ a_k^q > 0  \text{ and } q = 1,\dots, N_k^m $ &  Since no packet departed from the source node all packets in the queue will remain in the same queue position. Counters $a_k^q$ of status updates will be increased by one to account for the additional time-slot they will spend in the source node.\\ 
		\hline
		$a_k^q$ 		& $ a_k^q = -1  \text{ and }q = 1,\dots, N_k^m $ & Counters for application packets will not be incremented.\\
		\hline
		$-1$			& $ u_k^s = 0 \text{ and } w_k^a = 1 \text{ and } N_k^m \leq Q - 1 \text{and } q = N_k^m + 1 $ & An application packet arrival will be accommodated if there was at least one empty queue position during the $k$-th time-slot. \\
		\hline
		$1$ 			& $ u_k^s = 1\text{ and } N_k^m \leq Q - 1 \text{ and }q = N_k^m + 1 $ & A fresh status update will enter the queue before a new application packet. Given that fresh status updates are generated only when there exists at least one empty queue position there will always be place for the fresh status update. Application packets that find the queue full will be dropped.\\
		\hline
		$-1$, 			& $ u_k^s = 1\text{ and } w_k^a = 1\text{ and } N_k^m \leq Q-2 \text{and }  q = N_k^m + 2 $ & There will be enough queue positions to accommodate both a fresh status update and an application packet only if there were two empty queue slots during the $k$-th time-slot. \\
		\hline
		$0$, 			& $ q = N_k^m+2, \dots, Q $ & Counters $a_{k+1}^q$ will be set to zero for all empty queue positions.\\
		\hline
	\end{tabular}
\end{table*}

\subsubsection*{Transition cost and additive cost functions} With every state transition, according to control $u_k$, we associate a transition cost $g(x_k, u_k, w_k)$ which is defined as,
\begin{equation}
\label{eq:costFunction}
g(x_k, u_k, w_k) = \begin{cases}
G_{\Delta_{\max}}, & \text{if } x_k \in \Delta_{\max} \\
\Delta_{k+1}, & \text{otherwise},
\end{cases}
\end{equation}
where $w_k$ is the realization of random vector $W_k$ at the $k$-th time-slot and $G_{\Delta_{\max}}$ is a virtual cost associated with the employment of the expensive channel whenever $ x_k \in X_{\Delta_{\max}}$.
The value of $\Delta_{k+1}$ is completely determined by values $x_k$, $u_k$ and $w_k$, which are all known to the source node by the end of the $k$-th time-slot.

We are interested in minimizing the total cost accumulated over an infinite time horizon which is expressed as follows,
\begin{equation} 
\label{eq:cummulativeCostFunction}
J_{\pi}(x_0) = \underset{N \to \infty}{\lim} \underset{\underset{k=0,1,\dots}{W_k,}}{\mathop{\mathbb{E}}} \left\lbrace \sum_{k=0}^{N-1} \gamma^k g(x_k, u_k, w_k) | x_0\right\rbrace,  
\end{equation}
where $x_0$ is the initial state of the system, expectation $\mathbb{E}\lbrace \cdot \rbrace$ is taken with respect to the joint probability distribution of random variables $W_k$, $k=0, 1, \dots$ and $\gamma$ is a discount factor, i.e., $0 < \gamma < 1$, indicating that the importance of the induced cost decreases with time.
Finally, $\pi$ represents a policy, i.e., a sequence of functions $\pi = \{\mu_0, \mu_1, \dots \}$, where each function $\mu_k$ maps states to controls for the $k$-th stage. 
For a policy $\pi$ to belong to the set of all admissible policies $\Pi$, functions $\mu_k$ must satisfy the
constraint that for time-slot $k$ and state $x_k$ controls are selected exclusively from the set $U(x_k)$.

In order to minimize (\ref{eq:cummulativeCostFunction}), we must find an optimal policy $\pi^*$ that applies the appropriate control at each state. This is a non-trivial problem since control decisions cannot be viewed in isolation. 
One must balance the desire for low cost in the short-term with the risk of incurring high costs in the long run.
For example, a short sighted source node would avoid adding a fresh status update in a queue that already includes a status update. 
This is because the delay counter associated with the fresh status update will start incrementing immediately after its generation and this will have a negative impact on cost once the packet reaches the destination. 
However, this decision may lead to a queue filled with application packets and the AoI becoming equal to $\Delta_{\max}$, an event that will lead to the excessive penalty $G_{\Delta_{\max}}$. 

\section{Age Optimal Policies}
\label{sec:AgeOptimalPolicy}

The dynamic program presented in section~\ref{sec:SystemModel} is characterized by finite state, control, and probability spaces. 
Furthermore, transitions between states depend on $x_k$, $u_k$, and $w_k$ but not on their past values.
Additionally, the probability distribution of the random variables is invariant over time. 
Finally, the cost associated with a state transition is bounded and the cost function $J(\cdot)$ is additive over time.
Due to its structural properties the dynamic system at hand constitutes a Markov Decision Process (MDP)~\cite{B12} which is described by its state transition probabilities,

\begin{equation}
\label{eq:MdpTransitionProbabilities}
p_{ij}(u) = P \lbrace x_{k+1}=j|x_k=i,\ u_k=u \rbrace =  
	\sum_{(w_k^s, w_k^a) \in W_j} \hspace{-13px} P \lbrace W_k^s = w_k^s \rbrace P \lbrace W_k^a = w_k^a \rbrace		 
	\end{equation}

where, $x\in X,\ u\in U(x),\ (w_k^s, w_k^a) \in \{0,1\}^2$ and  $W_j = \{(w_k^s, w_k^a) \in \{0, 1\}^2: j = f(i, u, [w_k^s, w_k^a]^T)\}$.
From this point on we will utilize the MDP notation $p_{ij}(u)$ that presents the probability for the system to make a transition to state $j$ given that the system is in state $i$ and decision $u$ was made. 
	
For the MDP under consideration, given that $0<\gamma<1$, there exists an optimal stationary policy $\pi = \{\mu, \mu, \dots\}$, i.e., a policy that applies the same control function $\mu$ at all stages~\cite[Sec. 2.3]{B12}.
What is more, the control function $\mu$ will be independent of the initial state of the system and deterministic~\cite{B12}, i.e., each time the system is in state $i$, $\mu(i)$ applies the same control $u$. 
We will refer to a stationary policy $\pi = \{\mu, \mu, \dots \}$ as stationary policy $\mu$.
Our objective is to find a stationary policy $\mu^*$, from the set of all admissible stationary policies $\mathcal{M} \subseteq \Pi$, that minimizes the total cost in Equation~(\ref{eq:cummulativeCostFunction}), i.e., 
\begin{equation}
\mu^* = \arg \underset{\mu \in \mathcal{M}}{\min} J_{\mu}(i),\qquad \text{for all $i \in S$}.
\end{equation}
Let $J^*$ be the total cost attained when the optimal policy $\mu^*$ is used, then, for the MDP at hand, $J^*$ satisfies the Bellman equation,
\begin{align}
\label{eq:BellmanMDP}
J^*(i) &= \underset{u\in U(i)}{\min} \sum_{j = 1}^n p_{ij}(u) \left [g(i,u, j)+\gamma J^*(j) \right ],\text{ for all $i \in S$}, 
\end{align}
where $n$ is the cardinality of the state space. 
Equation~(\ref{eq:BellmanMDP}) describes a system of $n$ non-linear equations, the right hand side of which is a contraction, due to $\gamma < 1$, with a unique fixed point located at $J^*(i)$.
Due to the contraction property, one can derive both $J^*$ and $\mu^*$ via iterative methods. 

In this work we utilize the Optimistic Policy Iteration (OPI) algorithm~\cite{B12, sutton1998reinforcement, puterman2014markov} to approximate the optimal policy $\mu^*$ and the optimal infinite horizon cost $J^*$ for the problem under consideration. 
Part of the OPI algorithm is the Approximate Policy Evaluation (APE)~\cite{B12, sutton1998reinforcement, puterman2014markov} algorithm, used to evaluate the infinite horizon cost for the sequence of policies produced by the OPI in the process of approximating $\mu^*$.
APE is presented in Algorithm~\ref{alg:PolicyEvaluation}.
APE requires as input a stationary policy $\mu$ that maps each state $i \in X$ to a single control $u \in U(i)$ and returns an approximation of the infinite horizon cost $J_{\mu}$ for that policy. 
Optionally, if prior estimates for the values of $J_{\mu}$ exist, one may provide a $J_{\mu}$ in tabular form with preset cost values for each state $i \in X$, otherwise, APE will initialize arbitrarily the $J_{\mu}$.
APE will apply the transformation presented in the 5-th line of Algorithm~\ref{alg:PolicyEvaluation} to each state and will produce  $J_{\mu}'$ whose values are a closer estimate to the true values to the infinite horizon cost of policy $\mu$.
Formally, the values of $J_{\mu}$ will converge to the infinite horizon cost of policy $\mu$ only after an infinite number of repetitions. 
In practice, however, a finite number of repetitions is required for the algorithm to terminate and heuristically chosen values lead to an accurate calculation of $J_{\mu}$ as indicated by analysis and computational experience~\cite{B12}.
In Algorithm~\ref{alg:PolicyEvaluation} repetitions stop when $\underset{i\in X}{\max} |J_{\mu}'(i) - J_{\mu}(i)|$ becomes smaller than a predefined threshold $\epsilon$~\cite{sutton1998reinforcement}.
\begin{algorithm}
	\caption{Approximate Policy Evaluation}
	\label{alg:PolicyEvaluation}
	\begin{algorithmic}[1]
		\REQUIRE $\mu \in \mathcal{M}$
		\STATE Initialize $J_{\mu}(i) \in \mathbb{R},\; \forall i \in X$ arbitrarily if not given as input
		\STATE Initialize $\epsilon$ to a small value
		\REPEAT 
		\FORALL {$i \in X$}
		\STATE $J_{\mu}^{\prime} (i) \leftarrow \sum_{j=0}^{n} p_{ij}(\mu(i))[g(i,\mu(i),j) + \gamma J_{\mu}(j)]$
		\ENDFOR
		\STATE $D \leftarrow \underset{i\in X}{\max} |J_{\mu}'(i) - J_{\mu}(i)|$
		\STATE $J_{\mu} \leftarrow J_{\mu}'$
		\UNTIL $D < \epsilon$
		\STATE Return $J_{\mu}$
	\end{algorithmic}
\end{algorithm}

The OPI procedure is presented in Algorithm~\ref{alg:OptimisticPolicyIteration}.
OPI begins with arbitrarily initialized values for the policy $\mu$ and its infinite horizon cost $J$.
The values stored in tabular form will be updated iteratively and eventually will converge to $\mu^*$ and $J^*$.
The major operation of the OPI algorithm, besides calling APE, is presented in Line 5 and is called the \emph{policy improvement} step because its execution results in an improved policy $\mu'$, i.e., a policy that has a smaller infinite horizon cost compared to the previous policy $\mu$. 
Subsequently, APE is called with the improved policy $\mu'$ and $J$ as input. 
In this case $J$ is provided as a better initial guess for the infinite horizon cost for policy $\mu'$ compared to an arbitrarily set table of values and as a result the call to APE will terminate faster.
Upon termination APE will return an approximation for the infinite horizon cost of the improved policy $\mu'$ which will be subsequently used to derive an improved policy by the policy improvement step. 
According to the Bellman's optimality principle~\cite{B17, puterman2014markov}, unless policy $\mu_m$ is the optimal policy, the policy improvement step will always result in an improved policy, thus,
Algorithm~\ref{alg:OptimisticPolicyIteration} will terminate in case a policy improvement step does not result in an improved policy, i.e., $\mu' = \mu$.
Detailed analysis of the OPI and APE algorithms and their convergence properties can be found in~\cite{B12, B17, sutton1998reinforcement, puterman2014markov}. 
Finally, we note that the APE algorithm is also used to evaluate the infinite horizon cost for three heuristic policies that we will present in the next section.
\begin{algorithm}
\caption{Optimistic Policy Iteration}
\label{alg:OptimisticPolicyIteration}
\begin{algorithmic}[1]
	\STATE Initialize arbitrarily $J(i) \in \mathbb{R}$  and $\mu(i) \in U(i),\; \forall i\in X$.
	\REPEAT
	\STATE policy\_is\_stable $\leftarrow$ true
	\FORALL {$i \in X$}
	\STATE $\mu' (i) \leftarrow \arg \underset{u\in U(i)}{\min} [ \sum_{j=0}^{n} p_{ij}(\mu(i))(g(i,\mu(i),j) +\gamma J(j))]$
	\IF {$\mu'(i) \neq \mu(i)$} 
	\STATE policy\_is\_stable $\leftarrow$ false 
	\ENDIF 
	\ENDFOR
	\STATE $J \leftarrow$ APE($\mu'$, $J$)
	\STATE $\mu \leftarrow \mu'$	
	\UNTIL policy\_is\_stable
	\STATE Return $\mu \approx \mu^*$ and $J \approx J^*$
\end{algorithmic}
\end{algorithm}

\section{Results}
\label{sec:Results}
In this section we evaluate numerically the cost efficiency of the optimal policy $\mu^*$ for the system under consideration.
To provide insight into the structure of the optimal policy we introduce three heuristic policies and compare their cost efficiency with that of the optimal policy.

\begin{figure}[!htb]
	\centering
	\includegraphics[scale=0.9]{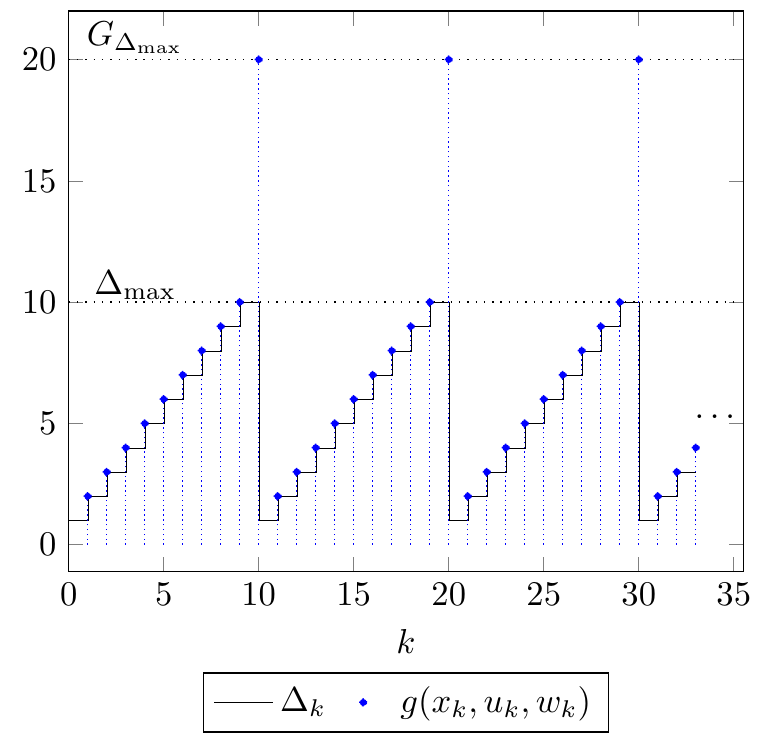}
	\vspace{-0.2in}
	\caption{AoI and transition cost for the never-sample policy when $\Delta_{\max}=10$ and $G_{\Delta_{\max}} = 20$. The transition cost is imposed on the source node at the end of each time slot.}
	\label{fig:neverSampleCostEvolution}
\end{figure}

The first heuristic policy is the \emph{zero-wait} policy, denoted with $\mu_{z}$, whereby the sensor will generate a status update either when the queue is empty or, mandatorily, when $x\in X_{\Delta_{\max}}$.
In both of these cases the status update will spend zero time waiting in queue.
The second heuristic policy is the \emph{max-sampling} rate policy, denoted with $\mu_{m}$, whereby the sensor will generate a status update in all states that this is permitted, i.e., in all states $x$ where $U(x)$ includes a control $u$ with $u^s = 1$ the max-sampling policy will select that specific control.
The third heuristic policy is the \emph{never-sample} policy, denoted with $\mu_{n}$, whereby the source node will never generate a status update unless this is mandatory, i.e., when $x\in X_{\Delta_{\max}}$.
The main characteristic of the never-sample policy is the periodicity of $\Delta$ and transition cost values.
More specifically, $\Delta$ will start with a value of one and will be incremented by one at each time-slot until, eventually, it becomes equal to $\Delta_{\max}$.
The cost for these state transitions, $g(x_k, u_k, w_k)$, is imposed \emph{at the end} of each time slot and its value is determined by the second branch of~(\ref{eq:costFunction}).
Once the threshold $\Delta_{\max}$ is reached, a status update will be transmitted through the expensive channel, resulting in a transition cost of $G_{\Delta_{\max}}$, and $\Delta$ will become equal to one again.
Fig.~\ref{fig:neverSampleCostEvolution} presents $\Delta_k$ and $g(x_k,u_k, w_k)$ for the never-sample policy when $\Delta_{\max}=10$ and $G_{\Delta_{\max}} = 20$.
The total cost over each period is given by, 
\begin{equation}
\label{eq:neverSamplePeriodicCost}
C_p = \sum_{c=2}^{\Delta_{\max}} c  + G_{\Delta_{\max}}= \frac{\Delta_{\max} (\Delta_{\max}+1)}{2}-1 + G_{\Delta_{\max}}.
\end{equation}
Never-sample policy exhibits the worst expected cost among all possible policies due to its complete lack of control over the status update process. 
In this work we also utilize its cost value as an indicator of how often the other three policies make use of the expensive channel.

\begin{table}[!t]
\caption{Basic Scenario Parameters}
\vspace{-0.2in}
\label{tbl:BasicScenarioParameters}
\centering
\begin{tabular}{|r|c|c|}
\hline
Description & Parameter & Value\\
\hline
\hline
Queue Size &	$Q$ & 4\\
AoI Threshold & $\Delta_{\max}$ & 10\\
Max. Retransmission Number & $r_{\max}$ & 4\\
Expensive channel cost &	$G_{\Delta_{\max}}$ & 100\\
Discount Factor &	$\gamma$ & 0.99\\
\hline
\end{tabular}
\end{table}

\begin{figure*}[!htb]
	\centering
	\subfloat[][$P_s = 0.8$]{
		\label{fig:Ps08}
		\includegraphics[scale=0.75]{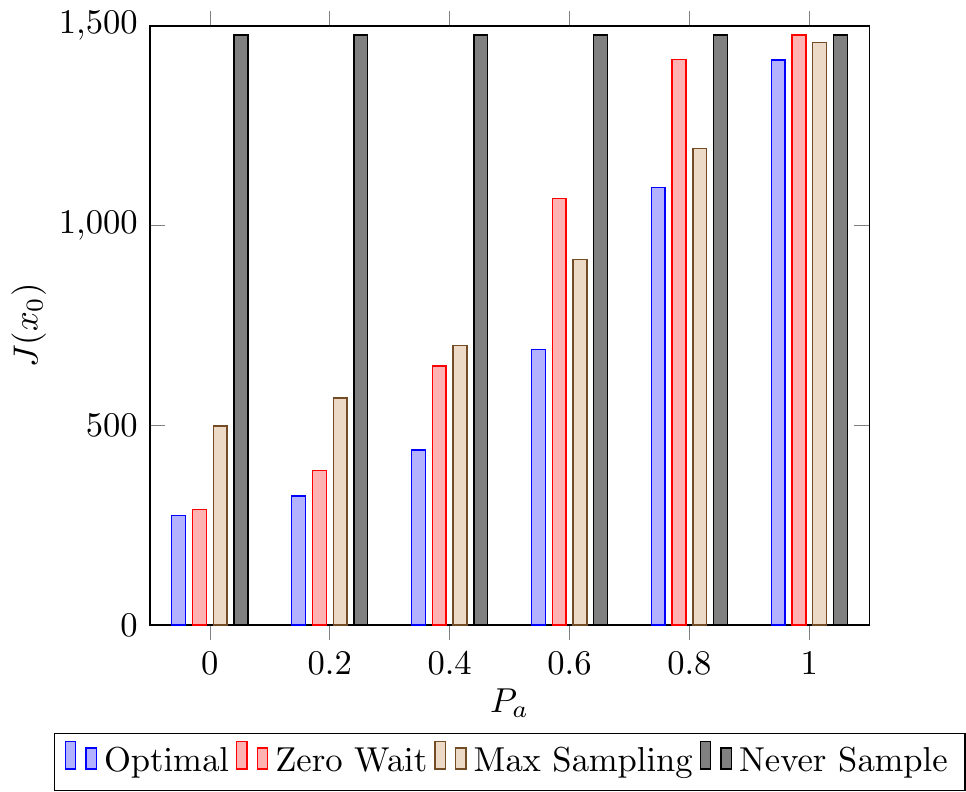} }
	\subfloat[][$P_s = 0.6$]{
		\label{fig:Ps06}%
		\includegraphics[scale=0.75]{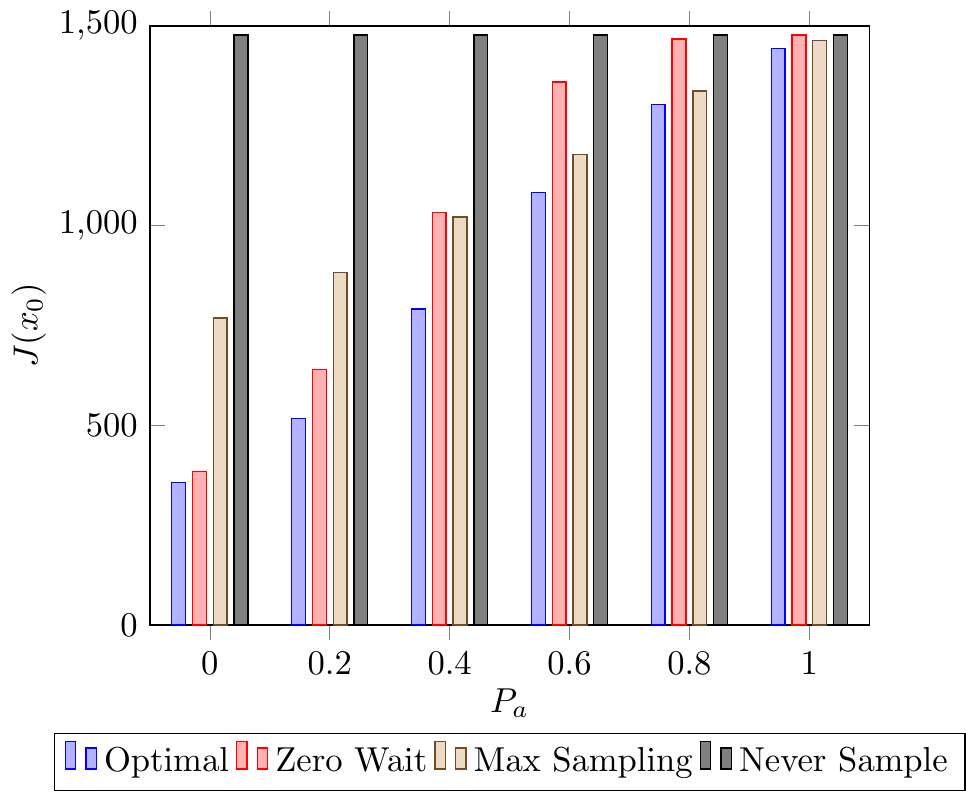}}\\
	\subfloat[][$P_s = 0.4$]{%
		\label{fig:Ps04}%
		\includegraphics[scale=0.9]{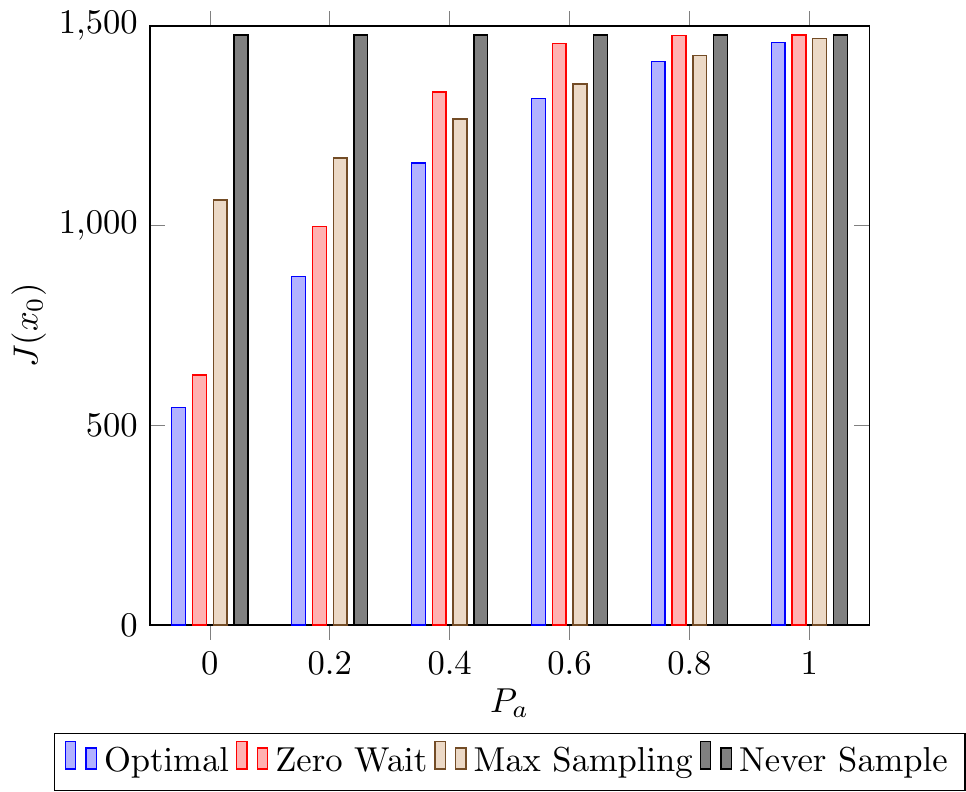}}
	\subfloat[][$P_s = 0.2$]{
		\label{fig:Ps02}
		\includegraphics[scale=0.9]{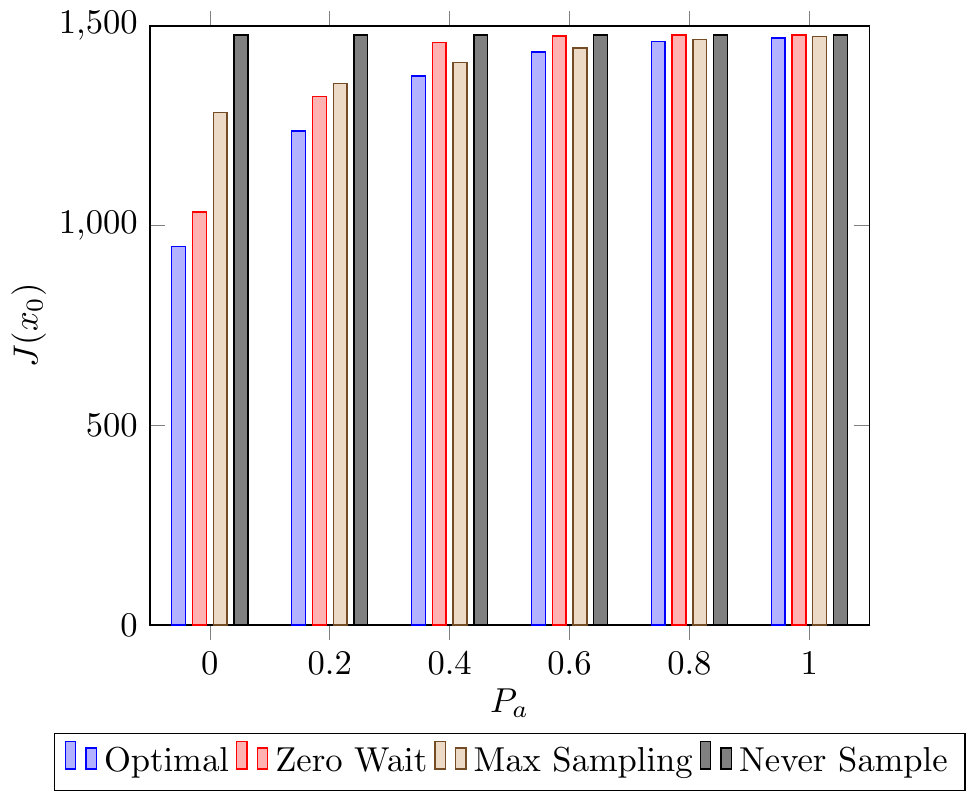}}%
	\vspace{-0.15in}
	\caption{Comparative plots for the optimal, zero-wait, max-sample and never-sample policies in terms of $J(x_0)$, i.e., the infinite horizon cost starting from an empty system with $\Delta_0 = 0$, for different values of the arrival probability for application packets $P_a$ and the probability for a successful transmission $P_s$.}
	\label{fig:basicScenario}
	\vspace{-20pt}
\end{figure*}

We consider the system of Fig.~\ref{fig:systemModel} configured with the set of parameter values presented in Table~\ref{tbl:BasicScenarioParameters}.
Let $x_0$ denote the initial state of the system, whereby the system is empty of packets and $\Delta_0 = 0$, then Fig.~\ref{fig:Ps08} presents the infinite horizon cost of all policies, i.e., $J_{\mu^*}, J_z, J_m, J_n$ for increasing values of the arrival probability $P_a$ and a successful transmission probability of $P_s = 0.8$.
In Fig.~\ref{fig:Ps08} and all subsequent figures we use $J(x_0)$ to refer to the cost associated with any policy. 
We note from Fig.~\ref{fig:Ps08} that when $P_a = 0$ or $P_a = 0.2$ the zero-wait policy is nearly optimal, as has  been already shown in the literature~\cite{yates2015lazy}.
This indicates that for a low value of $P_a$ the queue will often be empty of packets and a new status update will be generated frequently enough to avoid using the expensive channel.
On the other hand, the max-sampling policy performs poorly because it constantly fills the queue with status updates that consequently suffer long waiting times.
However, both the zero-wait and the max-sampling policies, as well as the optimal policy, achieve a much lower cost compared to the never-sample policy.
This result indicates that, unlike the never-sample policy, these policies successfully avoid high cost state transitions and especially the frequent use of the expensive channel.
This indication will become more concrete subsequently when we present results related to the frequency of usage of the expensive channel.
When $P_a = 0.4$ both zero-wait and max-sampling policies perform much worse than the optimal policy, a result that exhibits the inability of these policies to capture the trade-off between the arrival rates for status and application packets.
When $P_a$ is equal to $0.6$ or $0.8$ the max-sampling policy is a better approach to the optimal policy than the zero-wait policy.
This is due to the fact that application packets arrive at the queue with a high probability in each time-slot thus reducing the probability of an empty queue.
As a result the zero-wait policy will generate status updates less frequently and, consequently, will resort to the use of the expensive channel more often.
Finally, for $P_a = 1$, the optimal policy as well as all heuristic policies achieve similar costs. 
This indicates that the queue is always full with application packets and this causes the frequent use of the expensive channel by all policies in a way that resembles the operation of the never-sample policy.
For this latter policy, we see from Fig.~\ref{fig:Ps08} that its performance does not change with $P_a$ since it exclusively utilizes the expensive channel. 

Figures~\ref{fig:Ps06} to~\ref{fig:Ps02} present $J(x_0)$ for decreasing values of the probability to successfully transmit,  $P_s$.
With the exception of the never-sample policy, Figures~\ref{fig:Ps08} to~\ref{fig:Ps02} depict that, for a specific value of $P_a$, a decrement in $P_s$ results in an increased cost $J(x_0)$ for all policies.
As expected, unsuccessful packet transmissions increase the waiting time of all packets in the queue and often result in packet drops, which cause even larger values of $\Delta_k$, i.e., larger transition costs, and eventually lead to a more frequent use of the expensive channel.
The frequent use of the expensive channel is also indicated by Fig.~\ref{fig:Ps02} where all policies achieve a cost close to that of the never-sample policy even for relatively small values of $P_a$.
\begin{figure*}[!htb]
	\centering
	\subfloat[][$P_s = 0.8$]{
		\label{fig:steadyStateProbabilityToUseTheExpensiveChannel_08}
		\includegraphics[scale=0.75]{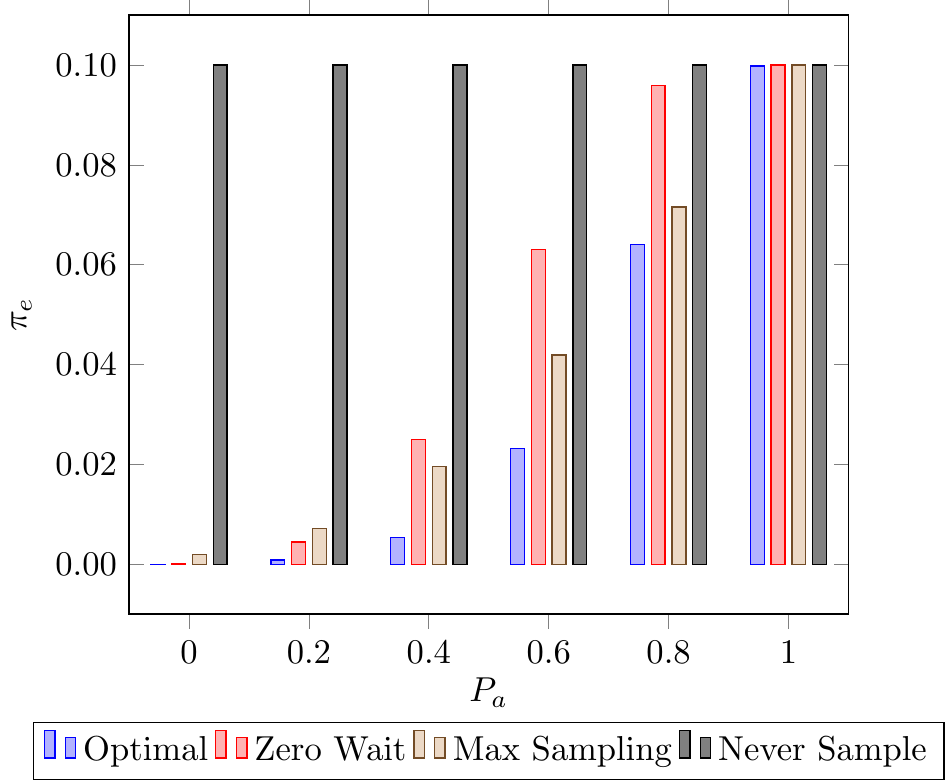} }
	\subfloat[][$P_s = 0.6$]{
		\label{fig:steadyStateProbabilityToUseTheExpensiveChannel_06}%
		\includegraphics[scale=0.75]{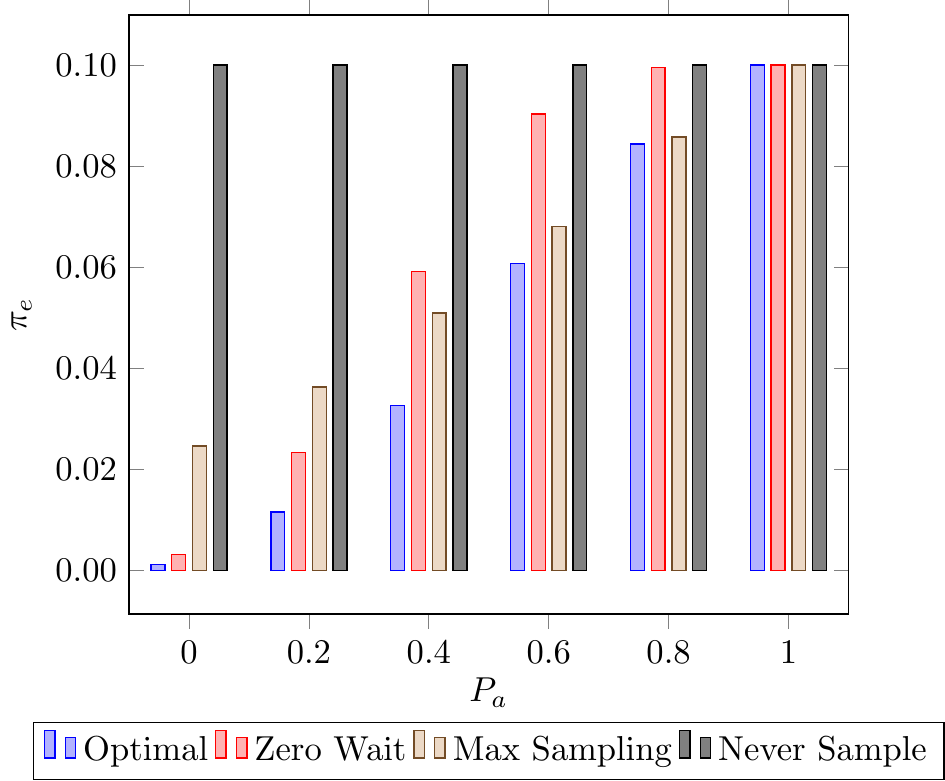}}\\
	\vspace{-10pt}
	\subfloat[][$P_s = 0.4$]{%
		\label{fig:steadyStateProbabilityToUseTheExpensiveChannel_04}%
		\includegraphics[scale=0.75]{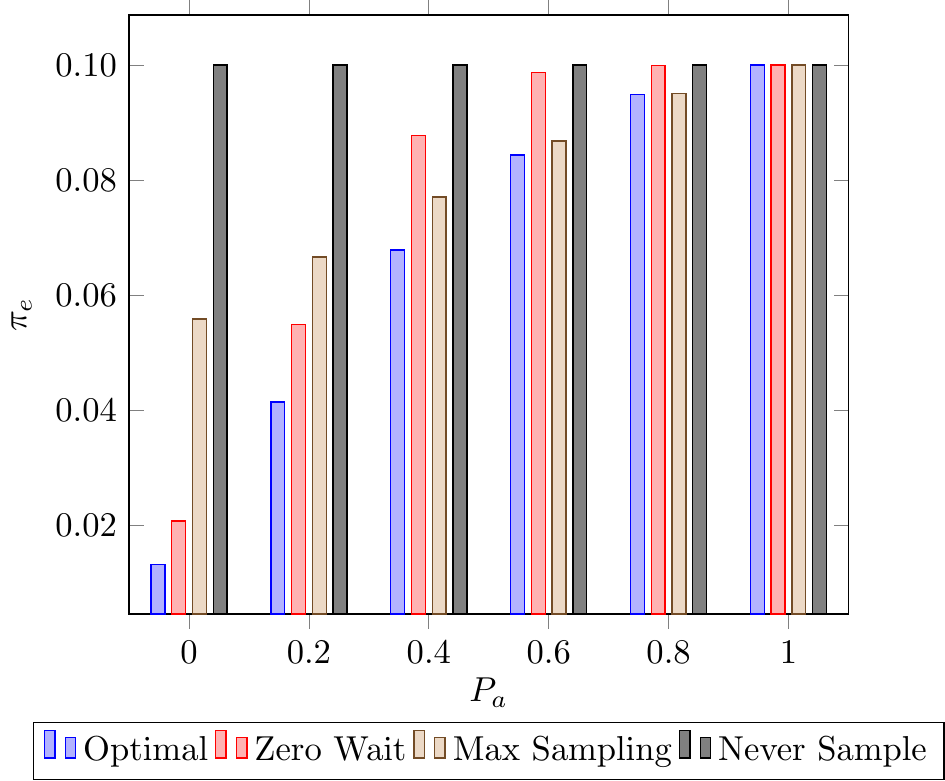}}
	\subfloat[][$P_s = 0.2$]{
		\label{fig:steadyStateProbabilityToUseTheExpensiveChannel_02}
		\includegraphics[scale=0.75]{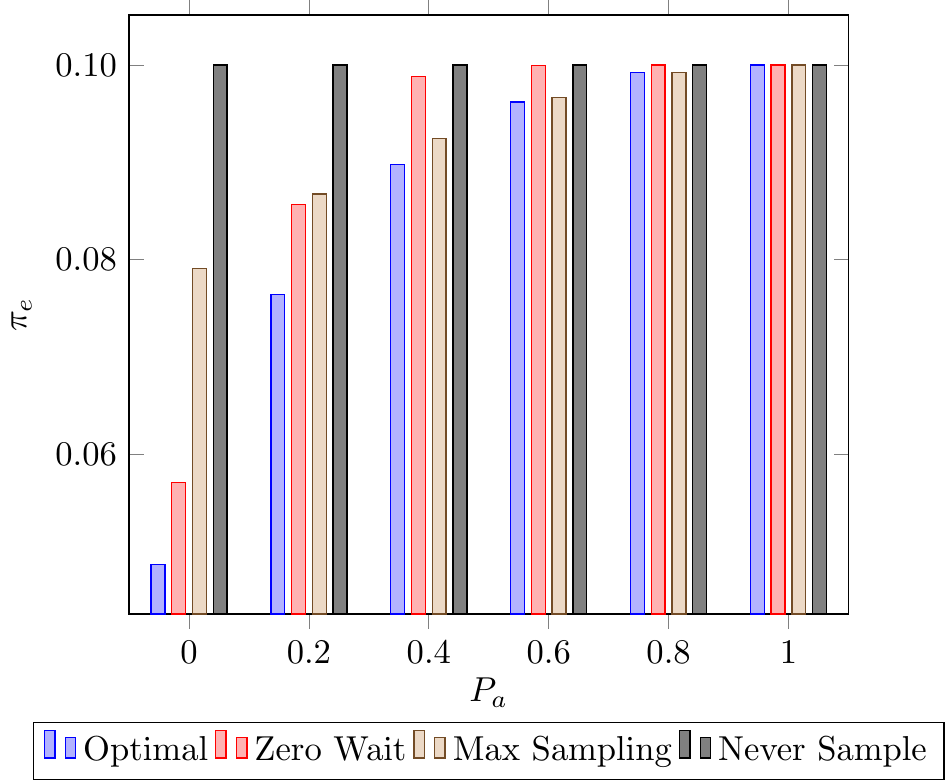}}%
	\vspace{-0.15in}
	\caption{Steady state probability to make use of the expensive channel, i.e., the sum of steady state probabilities for all states $x$ where $\Delta = \Delta_{\max}$. The never-sample policy is not amenable to the same probabilistic analysis as the other three policies due to its periodic character, yet the system will spend $10\%$ of its time using the expensive channel as is clearly shown in Fig.~\ref{fig:neverSampleCostEvolution}.}
	\label{fig:basicScenarioSteadyStateProbabilities}
\end{figure*}

To verify the assumption that the significant increase in $J(x_0)$ is due to the more frequent use of the expensive channel when $P_a$ increases or when $P_s$ decreases, we present in Fig.~\ref{fig:basicScenarioSteadyStateProbabilities} the aggregate steady state probability of the system being in a state that will result in using the expensive channel, i.e., the aggregate steady state probability to be in a state $x \in X_{\Delta_{\max}}$. 
We note that given $p_{ij}(u)$ for the MDP, as defined in~(\ref{eq:MdpTransitionProbabilities}), and the three stationary policies $\mu^*$, $\mu_z$ and $\mu_m$ one can derive the transition probability matrix $P$, for the resulting stochastic system as controlled by the provided policy.
For example, the elements of $P$ under the optimal policy are given by $P_{ij} = p_{ij}(\mu^*(i))$, for all $i, j \in X$.
To derive a steady state probability vector we focus on the recurrent class of states that includes the initial state $x_0$.  
Now let $P_r$ denote the transition probability matrix for this recurrent class of states, then we derive $\pi$, the steady state probability vector of $P_r$, as the normalized eigenvector of $P_r$ that corresponds to $P_r$'s eigenvalue $\lambda$ which is equal to one~\cite{gebali2015analysis}.
Finally, the aggregate steady state probability of the system to be in a state that will result in using the expensive channel
is given by, $\pi_{e} = \sum_{x \in X_{\Delta_{\max}}} \pi (x).$

The zero-sample policy is not amenable to the analysis presented above due to the periodic character of the resulting Markov process.
More specifically, the states of the resulting Markov process can be grouped in a finite number of disjoint subsets so that all transitions from one subset lead to the next~\cite{BT08}.
This is clearly shown in Fig.~\ref{fig:neverSampleCostEvolution} where a transition from a state with AoI equal to $\Delta$ will always lead to a state with AoI equal to $\Delta + 1$ unless $\Delta$ equals $\Delta_{\max}$, in which case a transition will lead to a state with $\Delta$ equal to one.
Therefore, by grouping states according to their AoI we can deduce the periodic character of the Markov process.
However, one can see from Fig.~\ref{fig:neverSampleCostEvolution} that the system will visit a state with AoI equal to $\Delta_{\max}$ once every $\Delta_{\max}$ transitions.
From this observation we can derive that it will spend $1/\Delta_{\max}$ of its time in states where the expensive channel is used.
For the scenarios in Fig.~\ref{fig:basicScenarioSteadyStateProbabilities} $\pi_e$ would be equal to 0.1. 
Figs.~\ref{fig:basicScenarioSteadyStateProbabilities}a-d exhibit that for large values of $P_a$ or low values of $P_s$ all policies behave the same way as the never-sample policy, i.e., they depend on the expensive channel.
Finally, we note that although all policies have the same steady state probability to use the expensive channel when $P_a = 1$,  as depicted in all cases of Fig.~\ref{fig:basicScenarioSteadyStateProbabilities}, they do not attain the same value of $J(x_0)$. 
This is due to the discount factor $\gamma$ being strictly less than one, which results in early transition costs having a larger impact on $J(x_0)$ compared to the transition costs for larger $k$ values.
More specifically, during the early stages, whereby the system begins with an empty queue, the optimal, zero-wait and max-sampling policies make better decisions compared to the never-sample policy and thus achieve relatively lower values of $J(x_0)$.

\begin{figure}[!htb]
	\centering
	\begin{minipage}[t]{.45\textwidth}
		\centering
		\includegraphics[scale=0.75]{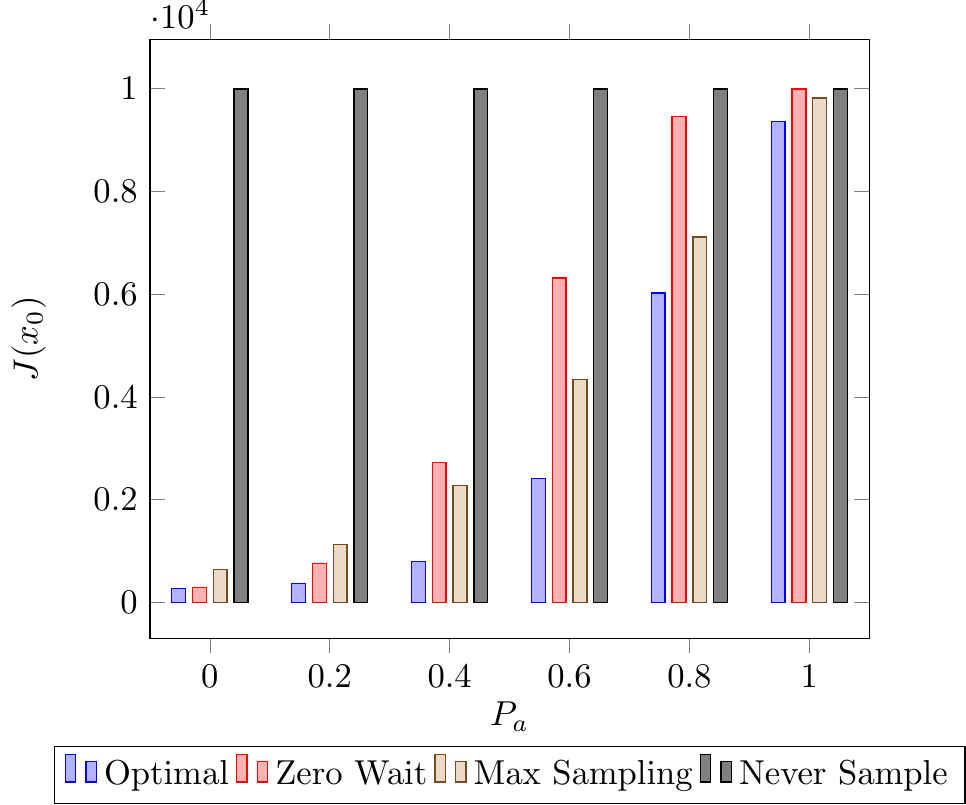}
		\vspace{-0.2in}
		\caption{Expected infinite horizon cost for all policies when the virtual cost $G_{\Delta_{\max}}$ associated with the use of the preemptive transmission mechanism is increased.}
		\label{fig:IncreaseMaxCost}
	\end{minipage}%
\hspace{20pt}
	\begin{minipage}[t]{.45\textwidth}
		\centering
		\includegraphics[scale=0.75]{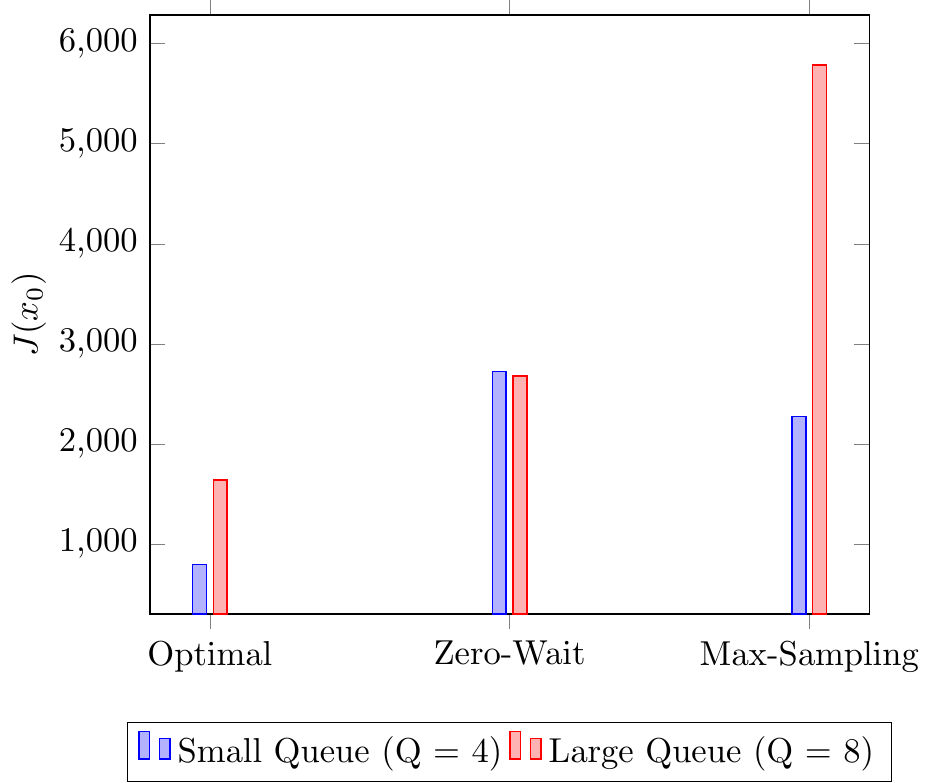}
		\vspace{-0.2in}
		\caption{Expected infinite horizon cost for three policies when the size of the queue doubles.}
		\label{fig:IncreasedQ}
	\end{minipage}
\end{figure}

Fig.~\ref{fig:IncreaseMaxCost} presents the impact of an increase of $G_{\Delta_{\max}}$ to the cost $J(x_0)$ when $P_s = 0.8$.
More specifically, we set $G_{\Delta_{\max}}=1000$ and note that the cost of the never-sample policy increases by an order of magnitude.
Comparing the results in Fig.~\ref{fig:IncreaseMaxCost} with those in Fig.~\ref{fig:Ps08} one can identify that for low values of $P_a$ cost $G_{\Delta_{\max}}$ has a small effect on the cost of all policies, with the exception of the never-sample policy.
This is justified by the fact that these policies resort infrequently to the use of the expensive channel when $P_a$ is low as has already be shown in Fig.~\ref{fig:basicScenarioSteadyStateProbabilities}.
On the other hand, for larger values of $P_a$ we observe a steep increment in cost which is due to the extensive use of the expensive channel. 		

Fig.~\ref{fig:IncreasedQ} presents the effect of an increase in the size of the queue on cost $J(x_0)$ for the optimal, zero-wait and max-sampling policies. 
More specifically, we increase the value of $Q$ from 4 to 8, while having $G_{\Delta_{\max}} = 1000$, $P_s = 0.8$ and $P_a = 0.4$.
Comparing the results of Fig.~\ref{fig:IncreasedQ} with the corresponding scenario of Fig.~\ref{fig:IncreaseMaxCost} one can see that the cost of the max-sampling policy has more than doubled due to the increased waiting times caused by the larger number of status updates that enter the queue. 
Similarly, the cost for the optimal policy has also increased significantly because the state space for the increased queue size scenario involves many states with a high cost expectancy, i.e., states with a large number of application packets that would incur increased waiting times and more frequent use of the expensive channel.
To avoid these states the controller has to make decisions that involve a more frequent generation of status updates so as to avoid using the expensive channel frequently.
However, these decisions involve higher values of $\Delta_k$ compared to the scenario with the same setup but a smaller queue, i.e., higher transition costs.
On the other hand, the cost for the zero-wait policy remains the same as that for a smaller queue size since the zero-wait policy takes control actions only when the queue is empty.
The rate with which the queue becomes empty depends on the values for $P_s$ and $P_a$ rather than the size of the queue, thus it was expected that the zero-wait policy would not be affected by an increment of the queue size.

\begin{figure}[!htb]
	\centering
	\includegraphics[scale=0.75]{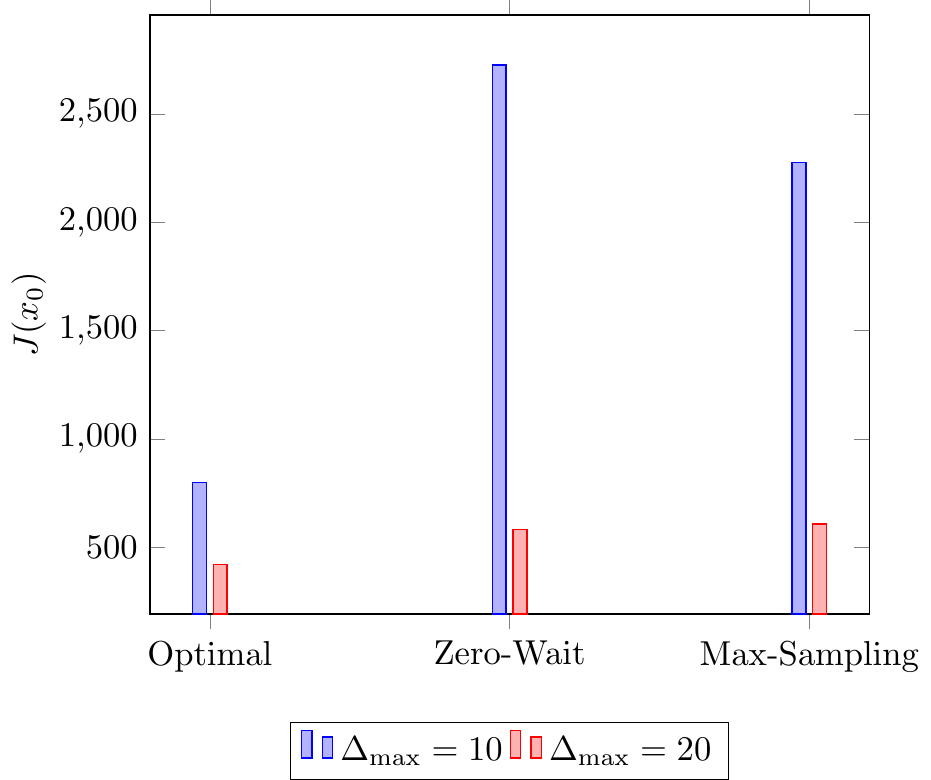}
	\vspace{-0.15in}
	\caption{Expected infinite horizon cost for three policies when the value of $\Delta_{\max}$ doubles.}
	\label{fig:IncreasedAoI}
	\label{fig:IncreasedDmax}
	\vspace{-15pt}
\end{figure}

Finally, Fig.~\ref{fig:IncreasedDmax} presents the effect of an increased $\Delta_{\max}$ value on cost $J(x_0)$.
More specifically, we increase the value of $\Delta_{\max}$ from 10 to 20, while having $G_{\Delta_{\max}} = 1000$, $P_s = 0.8$, $P_a = 0.4$ and $Q=4$.
Comparing the results in Fig.~\ref{fig:IncreasedDmax} with the corresponding scenario of Fig.~\ref{fig:IncreaseMaxCost} we see that by relaxing the constraint imposed by $\Delta_{\max}$, i.e., requiring less frequent status updates, the cost for all three policies is significantly reduced. 

\section{Conclusion}
\vspace{-0.2in}
\label{sec:conclusion}
In this work, we consider the problem of optimally controlling the generation of status updates for a communication system that serves the data traffic of two applications, one that is AoI sensitive and one that is not.
The data packets of both applications are stored in a single FIFO queue and they are transmitted via a wireless link to a destination node. 
We utilize the framework of Markov Decision Processes to derive optimal status update generation policies for a wide range of configurations and compare them against two baseline policies, the zero-wait policy and the max-sampling policy, where the latter policy generates status updates at a maximum rate.
The comparative results clearly exhibit that both baseline policies are suboptimal because they disregard the effect on AoI of the non-status update packet arrivals and the unsuccessful transmissions.
Furthermore, the results indicate the significant performance improvement resulting from the proposed problem formulation and the derived optimal policies.
However, a limitation of the current work is that the modeling framework of Markov Decision Processes is plagued with the curse of dimensionality which prohibits the efficient derivation of optimal policies for large scale systems due to the computational complexity involved in the process. 
As part of a future work we will apply approximate dynamic programming techniques on the current problem with the intention to derive near optimal policies in a computationally efficient way.

\appendices

\ifCLASSOPTIONcaptionsoff
\newpage
\fi

\bibliographystyle{IEEEtran}
\bibliography{bibliography}

\end{document}